# Plasticity of the Nb-rich μ-Co₇Nb₆ phase at room temperature and 600 °C

Authors

W. Luo [a,*], Z. Xie [a], P.-L. Sun [a], J. S. K.-L. Gibson [a, b], S. Korte-Kerzel [a,*]

Affiliation address

[a] Institute for Physical Metallurgy and Materials Physics, RWTH Aachen University, Kopernikusstraße 14, 52074 Aachen, Germany

[b] Department of Materials, University of Oxford, Parks Road, Oxford OX1 3PH, United Kingdom

[*] Corresponding authors

E-mail address: luo@imm.rwth-aachen.de (W. Luo); Korte-Kerzel@imm.rwth-aachen.de (S. Korte-Kerzel).

Abstract

The μ phase is a common precipitation phase in superalloys and it exists in a wide composition and temperature range. As such, we study the influences of composition and temperature on its plasticity by micropillar compression tests and transmission electron microscopy. The micropillars of the μ-Co₇Nb₆ phase deform plastically by basal slip at room temperature and 600 ºC. At room temperature, the Co-49Nb and Co-52Nb micropillars show high yield stresses and an abrupt large strain burst at the onset of yielding regardless of orientation, whereas the Co-54Nb micropillars oriented for basal slip yield at much lower stresses and show intermittent small strain bursts during plastic deformation. While the Co-49Nb micropillars deform by full dislocation slip on the basal plane at room temperature, the Co-54Nb micropillars deform by partial dislocation slip on the basal plane. At 600 ºC, the Co-49Nb micropillars oriented for basal slip show stable and continuous plasticity and their critical resolved shear stresses decrease dramatically. In contrast to the full dislocation slip at room temperature, the plastic deformation of the Co-49Nb micropillars occurs via partial dislocation slip on the basal plane at 600 ºC. Based on the geometric γ-surfaces for all potential basal slip planes, we explore where and why the glide of full and partial dislocations on the basal plane occurs in the μ-Co₇Nb₆ phase.

# 1    Introduction

The µ phase is a topologically close-packed (TCP) phase with an ideal stoichiometry $S_7L_6$, where the higher coordinated L atoms have a relatively larger atomic size than the icosahedrally coordinated S atoms. As shown in Fig. 1, the crystal structure of the µ phase is formed by alternate stacking of the $MgCu_2$-type Laves layer and the $Zr_4Al_3$ layer. A Laves layer consists of a kagomé layer and a triple layer [1]. The kagomé layer is commonly labelled with a capital Latin letter A, B or C. The large atoms at the top and the bottom of the triple layer are represented by a Greek letter α, β or γ, and the small atoms in the middle of the triple layer are represented by a lower case Latin letter a, b or c. Thus, the triple layer is denoted by αcβ, βaγ or γbα and the corresponding Lave layer is denoted by Aαcβ, Bβaγ or Cγbα. The $Zr_4Al_3$ layer consists of a kagomé layer, an layer of large atoms with CN14 coordination, and an layer of large atoms with CN15 coordination [2, 3].

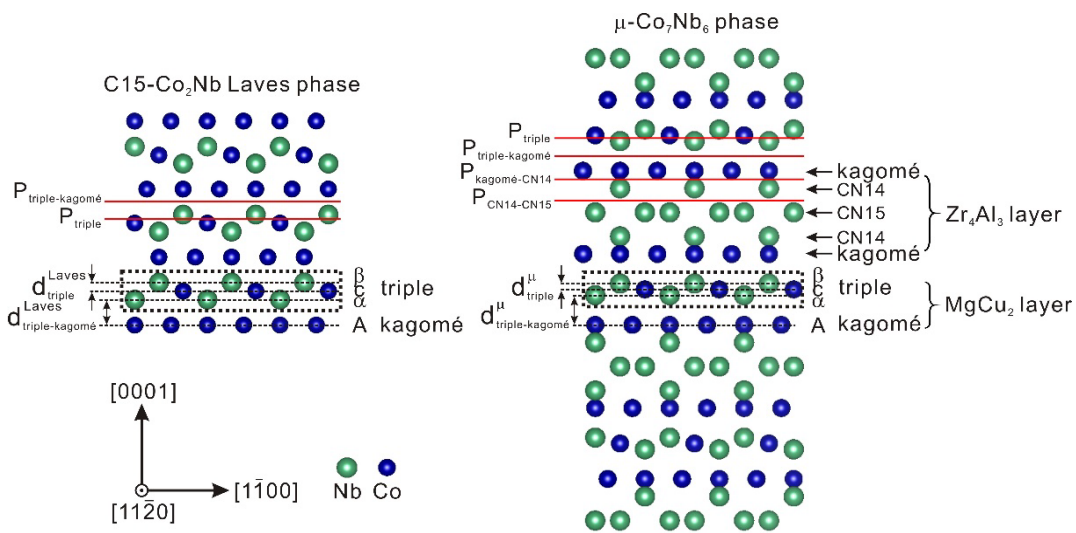

*Fig. 1 Atomic structure models of the C15-$Co_2Nb$ Laves phase and the µ-$Co_7Nb_6$ phase viewed along the $[11\bar{2}0]$ direction showing the stacking sequences and interplanar spacing. The $MgCu_2$–type Laves layer consists of an αcβ-type triple layer and an A-type kagomé layer. In the µ-$Co_7Nb_6$ phase it alternates with the $Zr_4Al_3$ layer which contains a kagomé layer and layers of large atoms having C14 and C15 coordinations. $d^i_{triple-kagome}$ and $d^i_{triple}$ (i = Laves or µ) represent the interplanar spacing between the triple layer and the kagomé layer, and the interplanar spacing within the triple layer, respectively.*

The µ phase is commonly observed in heavily alloyed superalloys [4-6] but the effect of the µ phase precipitation on the mechanical properties of superalloys is controversial and the magnitude of the effect remains unclear [7-16]. Due to the presence of exclusively tetrahedral interstices, the μ phase possesses a high packing density of atoms, and thus is usually considered to be hard and brittle. The μ phase can cause crack initiation and severe degradation of mechanical properties [7, 16]. The precipitation of the μ phase along grain boundaries can result in an embrittlement effect, leading to reductions in ambient temperature impact toughness [14] and tensile elongation [15]. On the contrary, it has been reported that although they are brittle, a small quantity of the μ phase precipitates does not affect the rupture mode and the ductility of the single crystal Ni-based superalloys during tensile, stress-rupture and impact tests [9, 13]. Furthermore, according to Yang et al. [8], the μ phase does not produce multiple internal cracks or interfacial decohesion in Ni-based superalloys during the room temperature tensile and high temperature stress rupture tests even when the content of the μ phase is high. Besides the embrittlement effect, it is believed that the μ phase precipitates can soften the Ni-based superalloys through the depletion of strengthening elements, leading to a reduction in stress rupture life at high temperature [17]. However, it has also been pointed out [9] that there is no conclusive observations to support the hypothesis. In addition to the embrittling and softening mechanisms [17], it has been reported that the precipitation of the μ phase could destroy the completeness of the raft structure [9, 11] and accelerate coarsening of the rafted structure [12], which leads to a detrimental effect of the high temperature creep properties of the Ni-based superalloys. In order to understand the role of the μ phase in the mechanical properties of superalloys, knowledge of the mechanical properties and deformation mechanism of the μ phase itself is required. However, due to the small size and brittle nature of the μ phase precipitates, our knowledge of the plasticity of the μ phase at both room and elevated temperature is very limited. Nanoindentation measurements of μ phase precipitates in the Ni-based superalloys at room temperature show that due to the close-packed structure and the high amounts of refractory elements, both the hardness and indentation modulus of the TCP phase are higher than those of the γ and γ' phases [18]. Recently, Schröders et al. [19, 20] reported that at room temperature the μ-$Fe_7Mo_6$ phase deforms plastically by dislocation slip predominately on the basal plane and that the faults observed by high resolution transmission electron microscopy are consistent with synchroshear. However, whether the basal slip in the μ phase generally occurs via the motion of synchroshear partial dislocations along the $\langle 1\bar{1}00 \rangle$ direction or by conventional slip of full dislocations along the $\langle 11\bar{2}0 \rangle$ direction is not clear yet. Although there is evidence suggesting plasticity of the μ phase at elevated temperatures [7, 8], no information on the deformation mechanism at elevated temperatures has been reported.

Overall, very little is known about the plasticity of the µ phase. Moreover, the µ phase is a complex intermetallic phase with potential mechanical anisotropy and a wide homogeneity range. The influences of orientation and composition on the plasticity of the µ phase also remain largely undiscovered. Due to the intrinsic brittleness of the µ phase, it is challenging to investigate its plasticity at ambient temperature by conventional mechanical testing. The difficulties in preparing flaw-less bulk samples of the µ phase and the brittle failure during conventional mechanical testing at ambient temperature can be avoided by micromechanical testing, such as micropillar compression [21, 22].

The aim of the present work is therefore to comprehensively study the plasticity of the µ phase with different orientations and compositions at both room and elevated temperatures, and to reveal the influence of orientation, composition, and temperature on the deformation behavior of the µ phase. We choose to work within the Co-Nb system, as in this system the congruently melting µ-$Co_7Nb_6$ phase exists in a broad temperature range up to 1424 ºC and a wide homogeneity range from $46.5 \pm 0.3$ to $56.1 \pm 0.4$ at.% Nb [23]. To investigate the activated slip systems and quantify the critical resolved shear stress (CRSS) of the µ-$Co_7Nb_6$ phase, we performed microcompression tests on micropillars of different sizes, compositions and orientations at both room temperature and 600 ºC. By means of transmission electron microscopy (TEM) we investigated the defects introduced during deformation and revealed the operating slip mechanisms. Finally, based on the geometric γ-surfaces for all potential basal slip planes, we rationalized the different dislocation types observed in the experiments.

## 2    Experimental methods

µ-$Co_7Nb_6$ phase alloys with nominal compositions of Co-49, 52 and 54 at.% Nb were prepared from Co (99.98 wt%) and Nb (99.9 wt%) by arc-melting in an argon atmosphere. To assure homogeneity, all samples were turned over and re-melted 5 times. The µ-$Co_7Nb_6$ phase alloys were heat treated at 1150 ºC for 500 h under high vacuum. The samples were cut into small discs with a thickness of 2 mm and a diameter of 8 mm, mechanically ground and polished using an OPU finish. The microstructures of the alloys were observed by optical microscopy (OM) (Leica DMR, Leica AG) and scanning electron microscopy (SEM) (Helios Nanolab 600i, FEI Inc.). Electron probe microanalysis (EPMA) (JXA-8530F, Jeol) measurements were carried out using pure Co and Nb as standards. Three line scans containing 33 spot analyses in total were measured for each sample. The compositions measured from the three line scans in different areas are consistent and the differences among them are within error bars. The average compositions of the nominally Co-49, 52 and 54 Nb (all in at.%) alloys obtained from EPMA measurements are $49.5 \pm 0.2$, $52.8 \pm 0.1$, and $53.7 \pm 0.2$ at.% Nb, respectively. The EPMA measurements show that the compositions of the alloys of the μ-Co$_7$Nb$_6$ phase are homogeneous and close to the nominal compositions. Crystallographic grain orientations were determined by electron backscatter diffraction (EBSD) (Hikari, EDAX Inc.).

*Table 1 Schmid factors of the full and partial dislocations associated with the primary basal slip systems of the micropillars of the μ-Co$_7$Nb$_6$ phase with various orientations under uniaxial compression.*

| Alloy | Temperature (ºC) | Loading axis/c-axis | Visualisation of loading direction | Schmid factor | | |
|---|---|---|---|---|---|---|
| | | | | full dislocation $\frac{1}{3}\langle11\bar{2}0\rangle$ | partial dislocations $\frac{1}{3}\langle1\bar{1}00\rangle$ | |
| Co-49Nb | 25 | 5º [0001] | 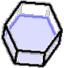 | 0.07 | 0.08 | 0.04 |
| | 600 | 12º [0001] | 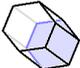 | 0.19 | 0.18 | 0.15 |
| | 600 | 34º [0001] | 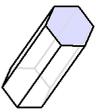 | 0.46 | 0.44 | 0.35 |
| | 25 | 39º [0001] | 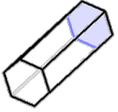 | 0.49 | 0.42 | 0.42 |
| | 25 | 47º [0001] | 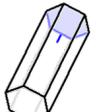 | 0.43 | 0.5 | 0.25 |
| | 25 | 88º [0001] | 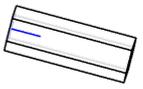 | 0.02 | 0.02 | 0.01 |
| Co-52Nb | 25 | 48º [0001] | 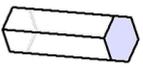 | 0.49 | 0.44 | 0.41 |
| | 25 | 50º [0001] | 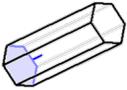 | 0.46 | 0.48 | 0.32 |
| | 25 | 46º [0001] | 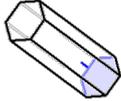 | 0.43 | 0.5 | 0.25 |
| Co-54Nb | 25 | 47º [0001] | 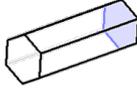 | 0.5 | 0.43 | 0.43 |
| | 25 | 84º [0001] | 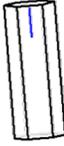 | 0.09 | 0.1 | 0.05 |

Cylindrical micropillars of the μ-Co$_7$Nb$_6$ phase with a top diameter ranging from 0.8 to 5 μm and an aspect ratio of about 2.5 were cut from the three μ-Co$_7$Nb$_6$ phase alloys by focused ion beam (FIB) (Helios Nanolab 600i, FEI Inc.) milling. A Ga$^+$ beam current of 2.5 nA at 30 kV was used for coarse milling. Final milling was carried out using 40 pA at 30 kV and 230 pA at 30 kV for preparations of small (0.8 μm in top diameter) and large (2, 3 and 5 μm in top diameter) micropillars, respectively. In Table 1, the orientations of the micropillars represented by the top views of the unit cells are denoted by the inclination angles between the loading axis and the *c*-axis. As different angles between the loading axis and the *c*-axis lead to different Schmid factors for basal slip, micropillars with specific orientations of nearly 0º, 45º and 90º inclined to the *c*-axis were prepared to study the activated basal and non-basal slip systems of the μ-Co$_7$Nb$_6$ phase. While basal slip is favourable when the *c*-axis is inclined to the loading axis by about 45°, non-basal slip might be activated when the angle between the loading axis and the *c*-axis is close to 0º or 90º due to the low Schmid factor for basal slip. Moreover, in materials with low stacking fault energy, different Schmid factors of the two partial dislocations leads to different resolved shear stresses acting on the partials, which can affect the partial dislocation separation distance, and thus the deformation mechanism [24, 25]. Therefore, the μ-Co$_7$Nb$_6$ phase micropillars oriented for basal slip were further divided into two groups: (1) in the micropillars oriented for $(0001)\langle1\bar{1}00\rangle$ partial dislocation slip, the primary $(0001)\langle1\bar{1}00\rangle$ slip system has the highest Schmid factor and the Schmid factor of the $\frac{1}{3}\langle1\bar{1}00\rangle$ leading partial dislocation is significantly higher than that of the $\frac{1}{3}\langle10\bar{1}0\rangle$ trailing partial dislocation, (2) in the micropillars oriented for $(0001)\langle11\bar{2}0\rangle$ full dislocation slip, the primary $(0001)\langle11\bar{2}0\rangle$ slip system has the highest Schmid factor and the Schmid factors of the two associated $\frac{1}{3}\langle 1\bar{1}00\rangle$ partial dislocations are similar. The Schmid factors of the full and partial dislocations associated with the primary slip systems are listed in Table 1.

The *in-situ* micropillar compression tests at room temperature and 600 °C were conducted inside a SEM (CLARA, Tescan GmbH) using an intrinsically load controlled indenter (InSEM, Nanomechanics Inc.). The Co-49, 52 and 54 Nb micropillars of the μ-$Co_7Nb_6$ phase were compressed at room temperature using a diamond flat punch of 10 μm diameter (Synton MDP AG) at a strain rate of about $10^{-3}$ $s^{-1}$ during stable loading. In addition, the Co-49Nb micropillars of the μ-$Co_7Nb_6$ phase of 2 μm diameter were compressed at 600 °C using a tungsten carbide flat punch of 10 μm diameter (Synton MDP AG) at a strain rate of about $10^{-3}$ $s^{-1}$ during stable loading. The indenter and the sample were heated independently. Before compression tests at 600 °C, the temperatures of the sample and the indenter were controlled by a thermocouple directly connected to the sample and a thermocouple placed next to the indenter shaft within the indenter tip heater, respectively. In order to eliminate thermal drift, the temperatures of the indenter and sample were tuned by using temperature measurements made with the indenter thermocouple [26]. The indenter heater was set to a constant heating power prior to indentation, and the indenter temperature variation was recorded during contact. If the temperatures of the indenter and the sample are matched, the indenter temperature will not significantly vary during contact. If the indenter temperature is observed to change, the indenter setpoint temperature is adjusted in the direction of the shift. During the compression tests, the force-displacement data was recorded. The engineering stress is calculated using the applied load and the cross-sectional area of the top surface. The engineering strain is calculated using the recorded displacement data of the indenter and the initial pillar height. After compression tests, the micropillars were imaged using SEM (Helios Nanolab 600i, FEI Inc.) and the slip traces on the pillar surface were analysed based on the grain orientations and *post-mortem* SEM images.

Specimens for transmission electron microscopy (TEM) analyses of the dislocation structures were prepared from the deformed micropillars of the μ-$Co_7Nb_6$ phase by FIB. The lamellae were cut with the foil normal directions perpendicular to the shear directions. TEM observations (JEM F200, Jeol and Tecnai F20, FEI) were carried out using double-tilt holders at an acceleration voltage of 200 kV.

To predict the energetically favourable slip systems in the μ-phase, a geometry-based method was applied in this work to generate a geometric γ-surface, as suggested by Pal et al. [27], by quantifying the variation of atomic volume during slip. The atomic volume of the μ-phase was calculated using the poly-disperse Voronoi tessellation algorithm implemented in OVITO [28], which considers the atomic radii of the elements. During the rigid body shift of a part of the crystal across a slip plane, the absolute deviation of the total atomic volume to the initial crystal was defined as the overlapped atomic volume. To generate the geometric γ-surface of the slip plane, the resulting overlapped atomic volumes were normalized by the number of unit cells in-plane and plotted on a two-dimensional surface.

## 3    Results

We present first the results of the compression tests of the Co-49Nb, Co-52Nb and Co-54Nb micropillars at room temperature, and then the results of the compression tests of the Co-49Nb micropillars at 600 °C, followed by the TEM analyses of the deformed micropillars.

### 3.1    Micropillar compression at room temperature

#### 3.1.1    Co-49Nb micropillars

The Co-49Nb micropillars of the μ-$Co_7Nb_6$ phase with 39º [0001] (Fig. 2 (b)) and 47º [0001] (Fig. 2 (c)) loading directions are oriented for $(0001)\langle 11\bar{2}0 \rangle$ and $(0001)\langle 1\bar{1}00 \rangle$ slip, respectively. They both show a large number of basal slip steps running through the entire micropillar. The Co-49Nb micropillars with 5º [0001] (Fig. 2 (a)) and 88º [0001] (Fig. 2 (d)) loading directions are oriented for non-basal slip. In contrast to the Co-49Nb micropillars oriented for basal slip, the 5º [0001] (Fig. 2 (a)) and 88º [0001] (Fig. 2 (d)) micropillars exhibit complex morphologies consisting of cracks and numerous fine basal slip traces. Neither prismatic nor pyramidal slip traces were observed. This indicates that non-basal slip is difficult to operate at room temperature and plasticity mainly occurs via basal slip in the Co-49Nb micropillars.

As shown in Fig. 2 (e), the engineering stress-strain curves of the Co-49Nb micropillars exhibit elastic deformation followed by a large strain burst at the onset of yielding regardless of pillar size and orientation. The yield stresses of the 39º [0001] micropillars of 0.8 and 2 μm diameter are 8 ± 2 and 5 ± 2 GPa, respectively. The yield stresses of the 47º [0001] micropillars of 0.8, 2 and 3 μm diameter are 8 ± 2, 5 ± 3 and 4.6 ± 0.9 GPa, respectively. The respective yield stresses of 5º [0001] micropillars of 0.8 and 2 μm diameter are 10 ± 1 and 6.4 ± 0.7 GPa. The respective yield stresses of 88º [0001] micropillars of 0.8 and 2 μm diameter are 10 ± 1 and 6 ± 1 GPa. In general, as pillar size increases, the yield stress decreases. The yield stresses of the micropillars of the same size are very close regardless of orientation.

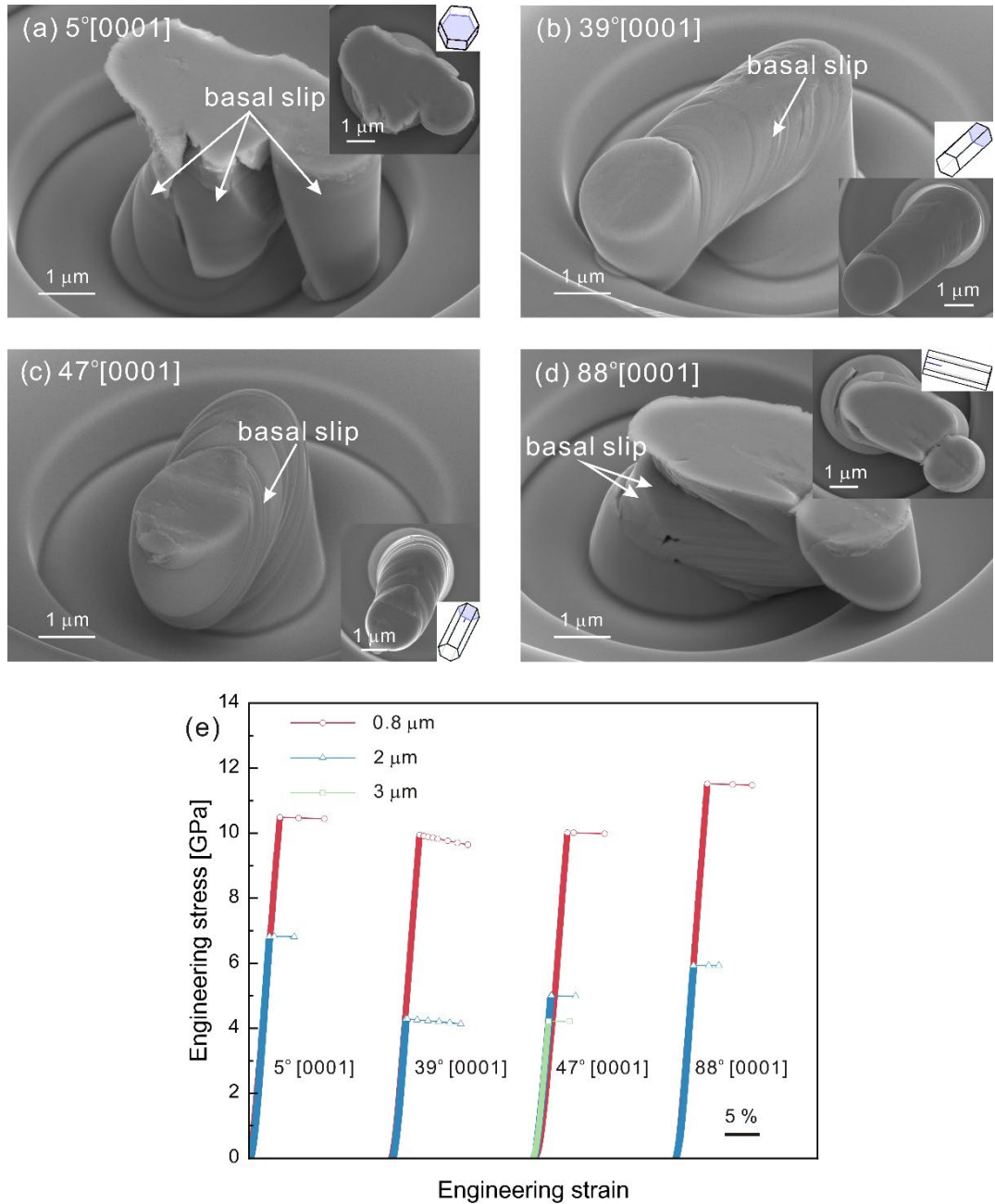

*Fig. 2 Representative post-mortem SEM images (side views after 45º tilt) of the Co-49Nb micropillars of the μ-Co₇Nb₆ phase with (a) 5º [0001], (b) 39º [0001], (c) 47º [0001] and (d) 88º [0001] orientations after compression at room temperature. Top views of the micropillars and unit cells showing the crystal orientations are inset. The basal slip traces are indicated by white arrows. (e) Representative engineering stress-strain curves of the Co-49Nb micropillars of the μ-Co₇Nb₆ phase with various sizes and orientations.*

### 3.1.2   Co-52Nb micropillars

All of the Co-52Nb micropillars of the μ-Co₇Nb₆ phase were oriented for basal slip. The Co-52Nb micropillars with 48º [0001] (Fig. 3 (a)) and 50º [0001] (Fig. 3 (b)) loading directions are oriented for $(0001)\langle 11\overline{2}0\rangle$ and $(0001)\langle 1\overline{1}00\rangle$ slip, respectively.

Similar to the Co-49Nb micropillars, the Co-52Nb micropillars in both orientations exhibit a high density of basal slip traces across the entire micropillar. Their stress-strain curves (Fig. 3 (c)) also show elastic deformation followed by a large strain burst at the onset of yielding. The yield stresses of the 48º [0001] micropillars decrease from 6 ± 1 to 5 ± 3 GPa, as pillar size increases from 0.8 to 2 μm. The yield stresses of the 50º [0001] micropillars decrease from 7 ± 2 to 5 ± 2 GPa, as pillar size increases from 0.8 to 2 μm. No pronounced influence of pillar orientation on the yield stresses of the Co-52Nb micropillars was found.

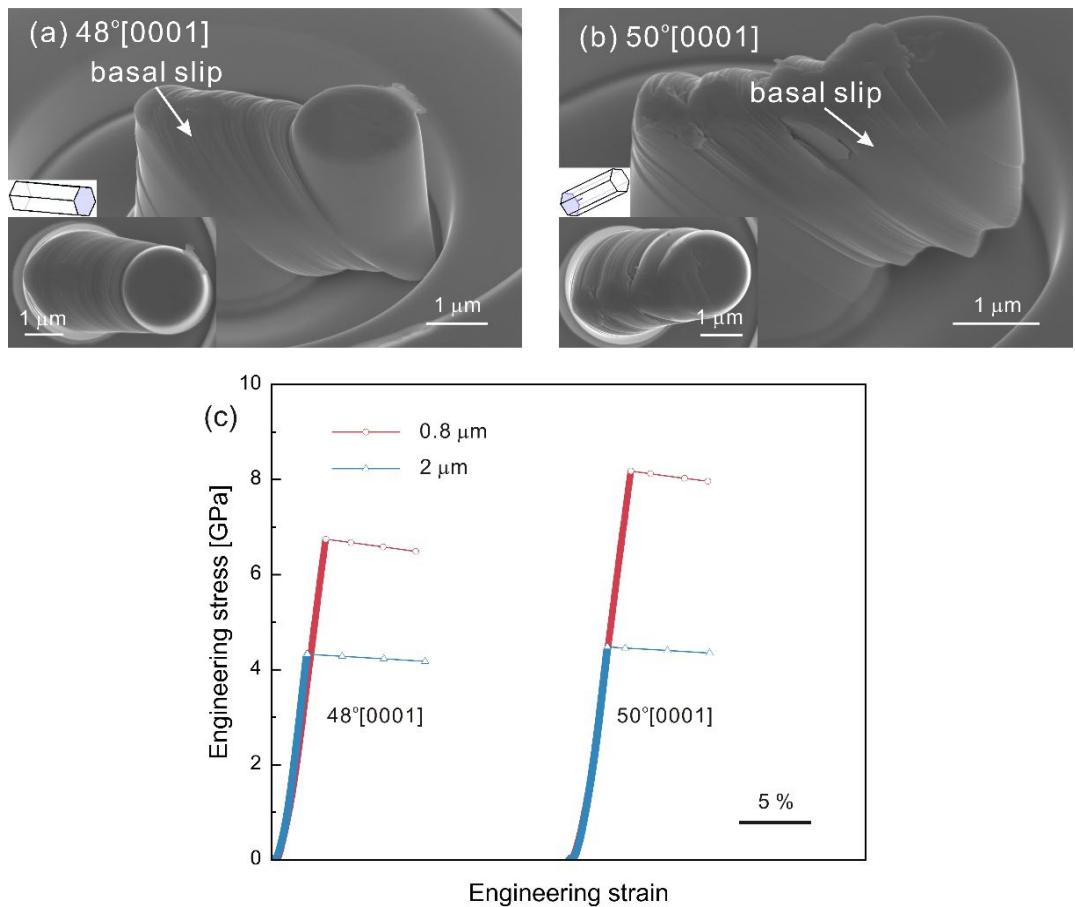

*Fig. 3 Representative post-mortem SEM images (side views after 45º tilt) of the Co-52Nb micropillars of the μ-Co₇Nb₆ phase with (a) 48º [0001] and (b) 50º [0001] orientations after compression at room temperature. Top views of the micropillars and unit cells showing the crystal orientations are inset. The basal slip traces are indicated by white arrows. (c) Representative engineering stress-strain curves of the Co-52Nb micropillars of the μ-Co₇Nb₆ phase with various sizes and orientations.*

### 3.1.3    Co-54Nb micropillars

The Co-54Nb micropillars of the μ-Co₇Nb₆ phase with 46º [0001] (Fig. 4 (a)) and 47º

[0001] (Fig. 4 (b)) orientations are oriented for $(0001)\langle 1\bar{1}00 \rangle$ and $(0001)\langle 11\bar{2}0 \rangle$ slip, respectively. In contrast to the Co-49Nb and Co-52Nb micropillars which were heavily deformed due to indenter overshooting, the plastic deformation of the Co-54Nb micropillars with 46º [0001] (Fig. 4 (a)) and 47º [0001] (Fig. 4 (b)) loading directions occurs in a more controllable manner so that loading can be stopped soon after yielding. The micropillars show clear basal slip traces without cracks and fracture. The Co-54Nb micropillars with 84º [0001] (Fig. 4 (c)) loading direction are oriented for non-basal slip. Although the basal plane is nearly parallel to the loading axis and the Schmid factor for basal slip is only 0.1 in the 84º [0001] (Fig. 4 (c)) micropillars, numerous slip traces parallel to the basal plane were observed on the surface. The 84º [0001] micropillars were severely deformed and fractured at the pillar base where gliding dislocations are blocked due to intense base constraints. No evidence of non-basal slip was found.

The representative engineering stress-strain curves of the Co-54 Nb micropillars are shown in Fig. 4 (d). In contrast to the high yield stresses and large strain bursts at the onset of yielding observed in the Co-49Nb and Co-52Nb micropillars, the Co-54Nb micropillars oriented for basal slip yield at much lower stresses and show intermittent small strain bursts during plastic deformation. The stress where the first strain burst occurs is taken as the yield stress. The yield stresses of the 46º [0001] micropillars of 0.8, 2 and 5 μm diameter are $2.0 \pm 0.5$, $1.4 \pm 0.3$ and $0.9 \pm 0.2$ GPa, respectively. The yield stresses of the 47º [0001] micropillars of 0.8, 2 and 3 μm diameter are $2.3 \pm 0.7$, $1.2 \pm 0.4$ and $1.2 \pm 0.4$ GPa, respectively. However, the Co-54Nb micropillars with 84º [0001] orientation, which have a low Schmid factor for basal slip, show linear-elastic deformation followed by abrupt failure at relatively high stresses. The respective yield stresses of the 84º [0001] micropillars of 0.8 and 2 μm diameter are $7.8 \pm 0.4$ and $5.1 \pm 0.4$ GPa, which are much higher than those of the micropillars oriented for basal slip.

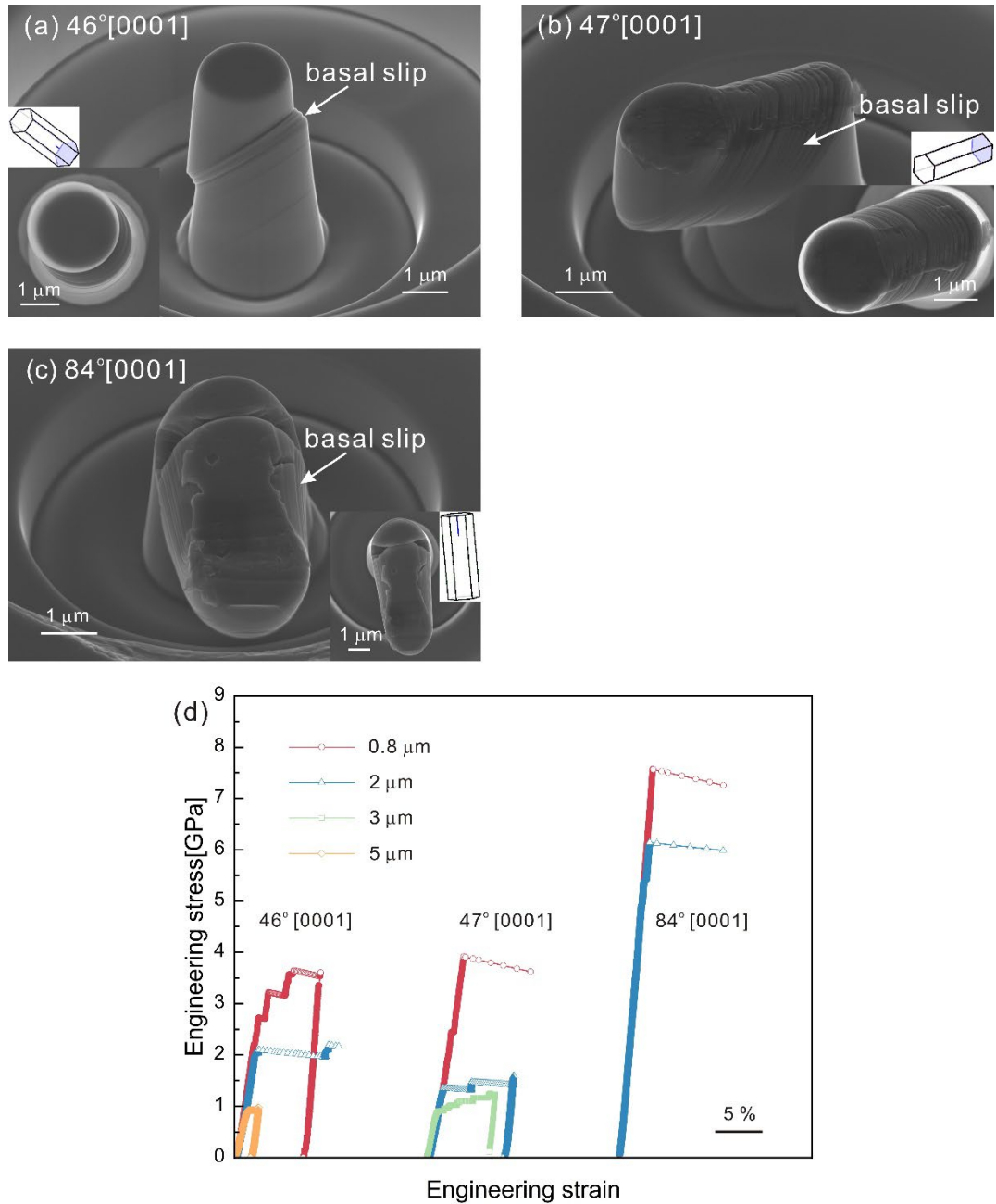

*Fig. 4 Representative post-mortem SEM images (side views after 45º tilt) of the Co-54Nb micropillars of the μ-Co₇Nb₆ phase with (a) 46º [0001] and (b) 47º [0001], and (c) 84º [0001] orientations after compression at room temperature. Top views of the micropillars and unit cells showing the crystal orientations are inset. The basal slip traces are indicated by white arrows. (d) Representative engineering stress-strain curves of the Co-54Nb micropillars of the μ-Co₇Nb₆ phase with various sizes and orientations.*

## 3.2 Micropillar compression at 600 ºC

While the Co-49Nb micropillars of the μ-Co$_7$Nb$_6$ phase with 12º [0001] (Fig. 5 (a)) loading direction are oriented for non-basal slip due to the low Schmid factor for basal slip, the Co-49Nb micropillars of the μ-Co$_7$Nb$_6$ phase with 34º [0001] (Fig. 5 (b)) loading direction are oriented for basal slip. After compression at 600 ºC, the 12º [0001] oriented Co-49Nb micropillars (Fig. 5 (a)) show cracks accompanied by complex traces at the pillar top and basal slip traces transecting the middle of the micropillar. In contrast, the 34º [0001] oriented Co-49Nb micropillars exhibit only basal slip traces at the pillar top without non-basal traces and cracks.

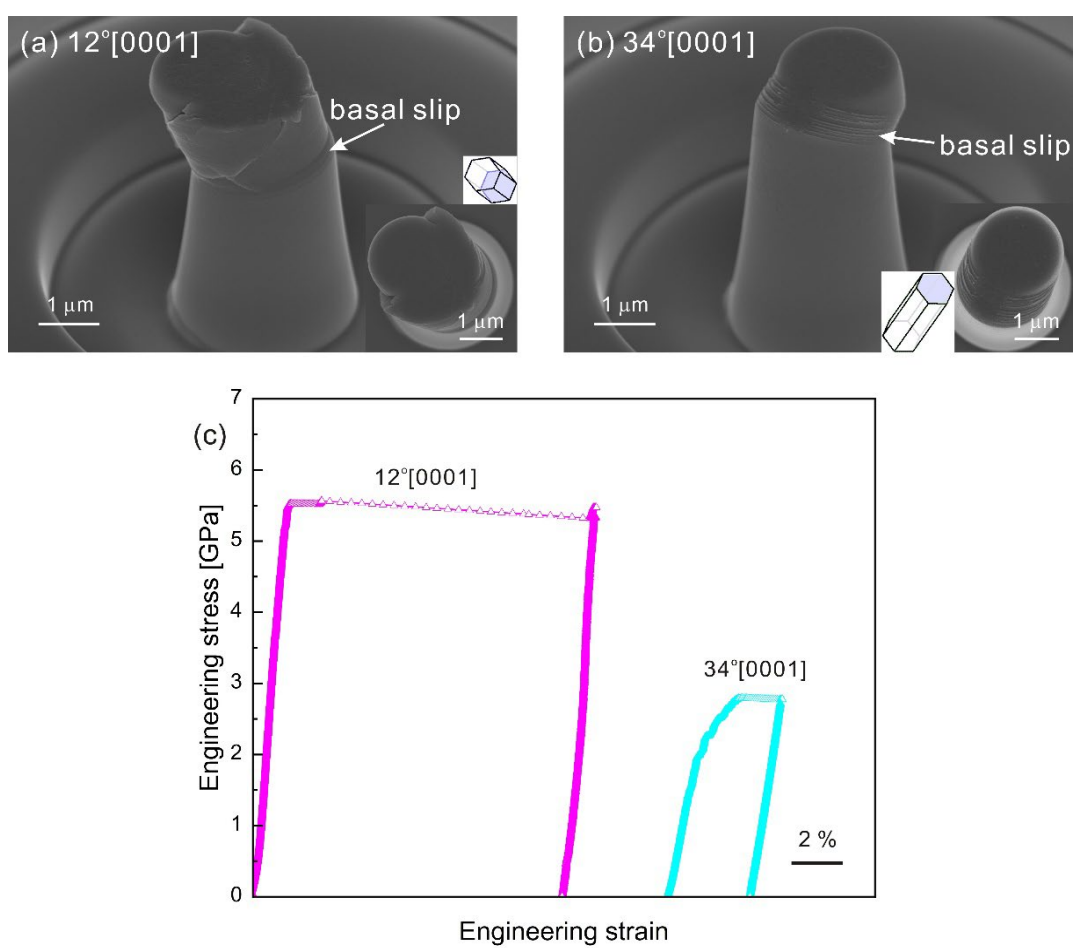

*Fig. 5 Representative post-mortem SEM images (side views after 45º tilt) of the Co-49Nb micropillars of the μ-Co$_7$Nb$_6$ phase of 2 μm diameter with (a) 12º [0001] and (b) 34º [0001] orientations after compression at 600 ºC. Top views of the micropillars and unit cells showing the crystal orientations are inset. The basal slip traces are indicated by white arrows. (c) Representative engineering stress-strain curves of the Co-49Nb micropillars of the μ-Co$_7$Nb$_6$ phase with various sizes and orientations at 600 ºC.*

A representative stress-strain curve of the 12º [0001] oriented Co-49Nb micropillar of 2 μm diameter (Fig. 5 (c)) at 600 °C shows linear elastic deformation followed by a limited amount of plastic deformation and an abrupt large strain burst. The average yield stress of the 12º [0001] micropillars of 2 μm diameter is $5.0 \pm 0.5$ GPa. In contrast, at 600 °C the 34º [0001] micropillars of 2 μm diameter oriented for basal slip yield at a much lower stress of $1.9 \pm 0.3$ GPa and show stable plastic deformation with intermittent small strain bursts after yielding.

### 3.3 TEM investigations

### 3.3.1 Co-49Nb micropillars deformed at room temperature

In order to study the deformation mechanisms of basal slip in the μ-$Co_7Nb_6$ phase beyond the visible slip traces at the surface and the associated yield stresses and characteristics, TEM investigations were performed on the deformed Co-49Nb micropillars oriented for basal slip.

Figs. 6 (a-d) show the TEM bright-field images of a Co-49Nb micropillar with 47º[0001] loading direction (oriented for $(0001)\langle 1\bar{1}00\rangle$ slip) taken under different imaging conditions. As shown in Fig. 6 (a), highly strained slip bands, individual dislocations, and a few widely extended stacking faults parallel to the basal plane were observed when imaged along the $[2\bar{1}\bar{1}0]$ zone axis using $g = 02\bar{2}1$. As the arrays of dislocations are lying on the basal slip bands and induce strain contrast, the dislocations are likely induced by plastic deformation. All the defects are out of contrast under $g = 000\bar{3}$ (Fig. 6 (b)), which confirms that the dislocations and stacking faults lie on the basal plane. When imaged along the $[10\,\overline{10}\,0\,\bar{1}]$ zone axis, the dislocations marked by a red ellipse are visible using $g = 01\overline{1}\overline{1}0$ (Fig. 6 (c)) but they are out of contrast using $g = 1\,0\,\bar{1}\,10$ (Fig. 6 (d)). No contrast of stacking faults was observed between the dislocations. Therefore, the dislocations marked by the red ellipse are $\frac{1}{3}[1\bar{2}10]$ full dislocations. Although partial dislocation slip on the basal plane is favorable in this orientation, no evidence of dislocation dissociation was found.

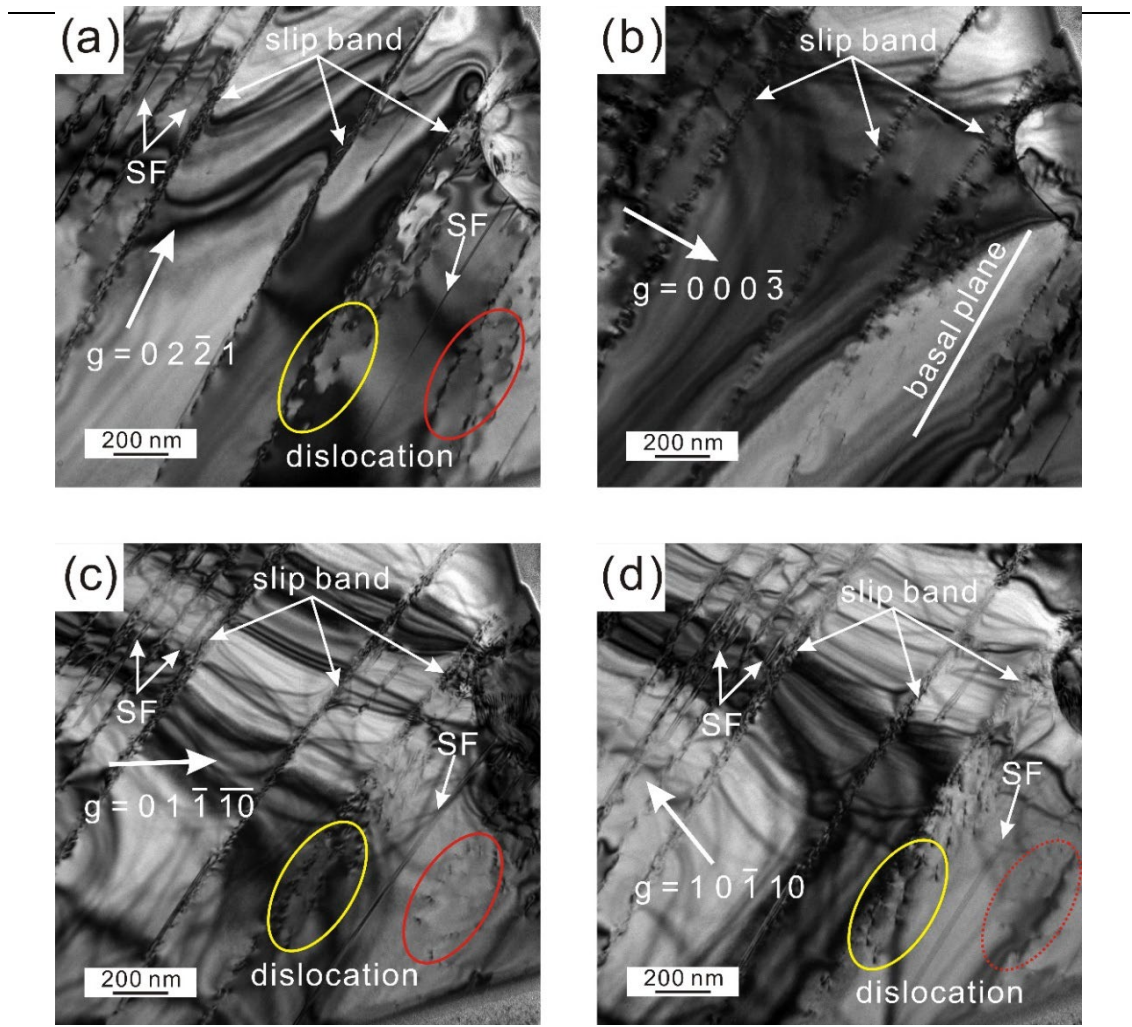

*Fig. 6 TEM bright-field images of a Co-49Nb micropillar with 47º[0001] loading direction (oriented for $(0001)\langle 1\bar{1}00\rangle$ slip) taken along $[2\bar{1}\bar{1}0]$ zone axis using (a) $g = 02\bar{2}1$ and (b) $g = 000\bar{3}$, and along $[10\bar{1}0\,0\,\bar{1}]$ zone axis using (c) $g = 01\bar{1}\overline{10}$, and (d) $g = 1\,0\,\bar{1}\,10$.*

Figs. 7 (a-d) show TEM bright-field images of a Co-49Nb micropillar with 39º[0001] loading direction (oriented for $(0001)\langle 11\bar{2}0\rangle$ slip) taken under different two-beam conditions. When imaged along the $[0\bar{1}10]$ zone axis, a high density of dislocations was observed using $g = 2\bar{1}\overline{10}$ (Fig. 7 (a)) and they are out of contrast with $g = 0003$ (Fig. 7 (b)), which indicates that the dislocations are on the basal plane. According to the invisibility criterion, i.e. $g \cdot b = 0$, the dislocations marked by red ellipses and yellow ellipses (Figs. 7 (c) and (d)) are $\frac{1}{3}[11\bar{2}0]$, and $\frac{1}{3}[1\bar{2}10]$ full dislocations, respectively. The presence of dislocations with different Burgers vectors suggests that multiple $(0001)\langle11\bar{2}0\rangle$ slip systems are activated, possibly due to friction between the pillar top and the flat punch or small misalignments. Therefore, the TEM analyses show that both the Co-49Nb micropillars oriented for $(0001)\langle1\bar{1}00\rangle$ and $(0001)\langle11\bar{2}0\rangle$ slip deform by $\frac{1}{3}\langle11\bar{2}0\rangle$ full dislocation slip on the basal plane at room temperature.

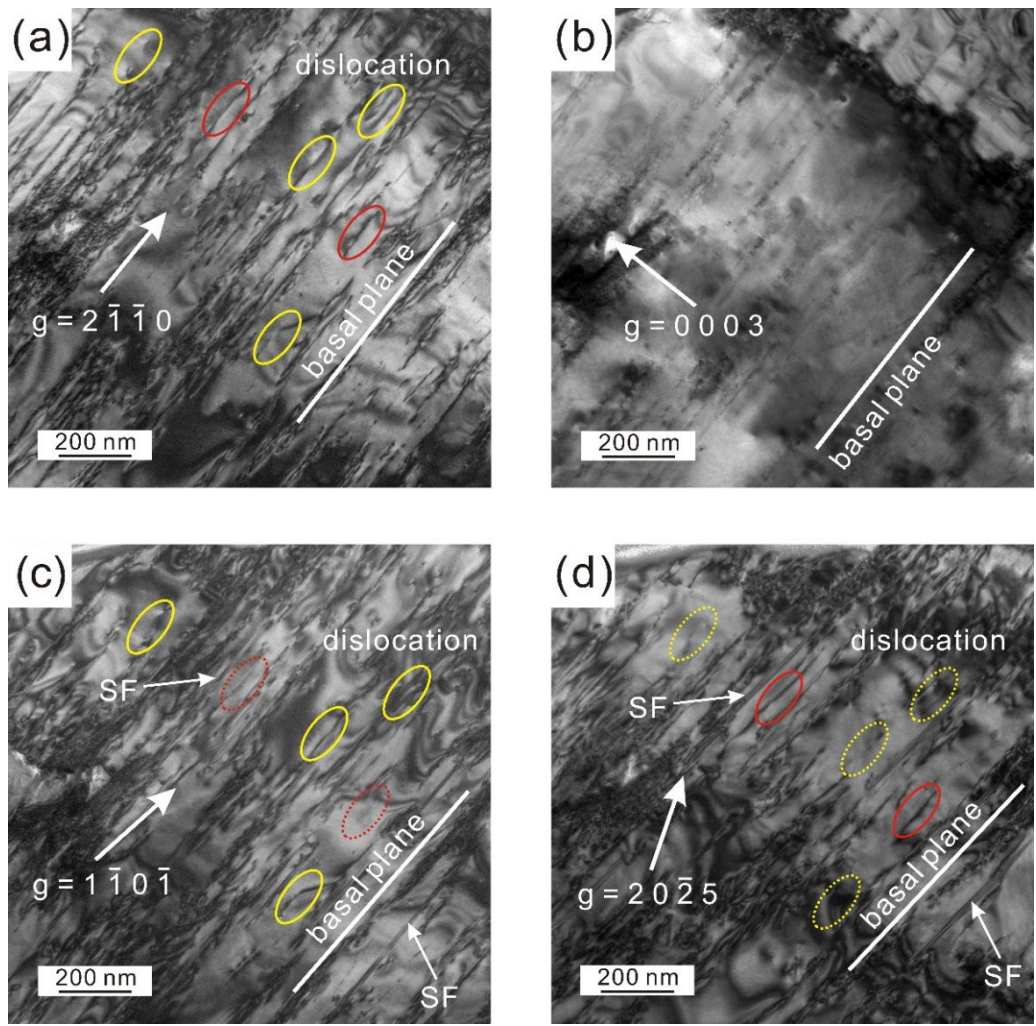

*Fig. 7 TEM bright-field images of a Co-49Nb micropillar with 39º[0001] loading direction (oriented for $(0001)\langle11\bar{2}0\rangle$ slip) taken along $[0\bar{1}10]$ zone axis using (a) $g = 2\bar{1}\bar{1}0$ and (b) $g = 0003$, and along $[\bar{1}\bar{1}20]$ zone axis using (c) $g = 1\bar{1}0\bar{1}$, and along $[1\bar{2}10]$ zone axis using (d) $g = 20\bar{2}5$.*

### 3.3.2 Co-54Nb micropillars deformed at room temperature

Figs. 8 (a-d) show TEM bright-field images of a Co-54Nb micropillar with 46º[0001] loading direction (oriented for $(0001)\langle 1\overline{1}00\rangle$ slip) taken under different two-beam conditions. When imaged along the $[\overline{1}010]$ zone axis, dislocations were observed using $g = 1\overline{2}10$ (Fig. 8 (a)) and they are out of contrast using $g = 00012$ (Fig. 8 (b)). When imaged along the $[\overline{10}\,0\,10\,\overline{1}]$ zone axis, contrast of stacking faults between the partial dislocations was observed on the basal plane using $g = 0\,\overline{1}\,110$ (Fig. 8 (c)). The stacking faults are out of contrast with $g = 2\overline{4}20$ (Fig. 8 (d)). The pile-ups of partial dislocations bounding stacking faults indicate repeated operation of partial dislocation sources on the basal plane.

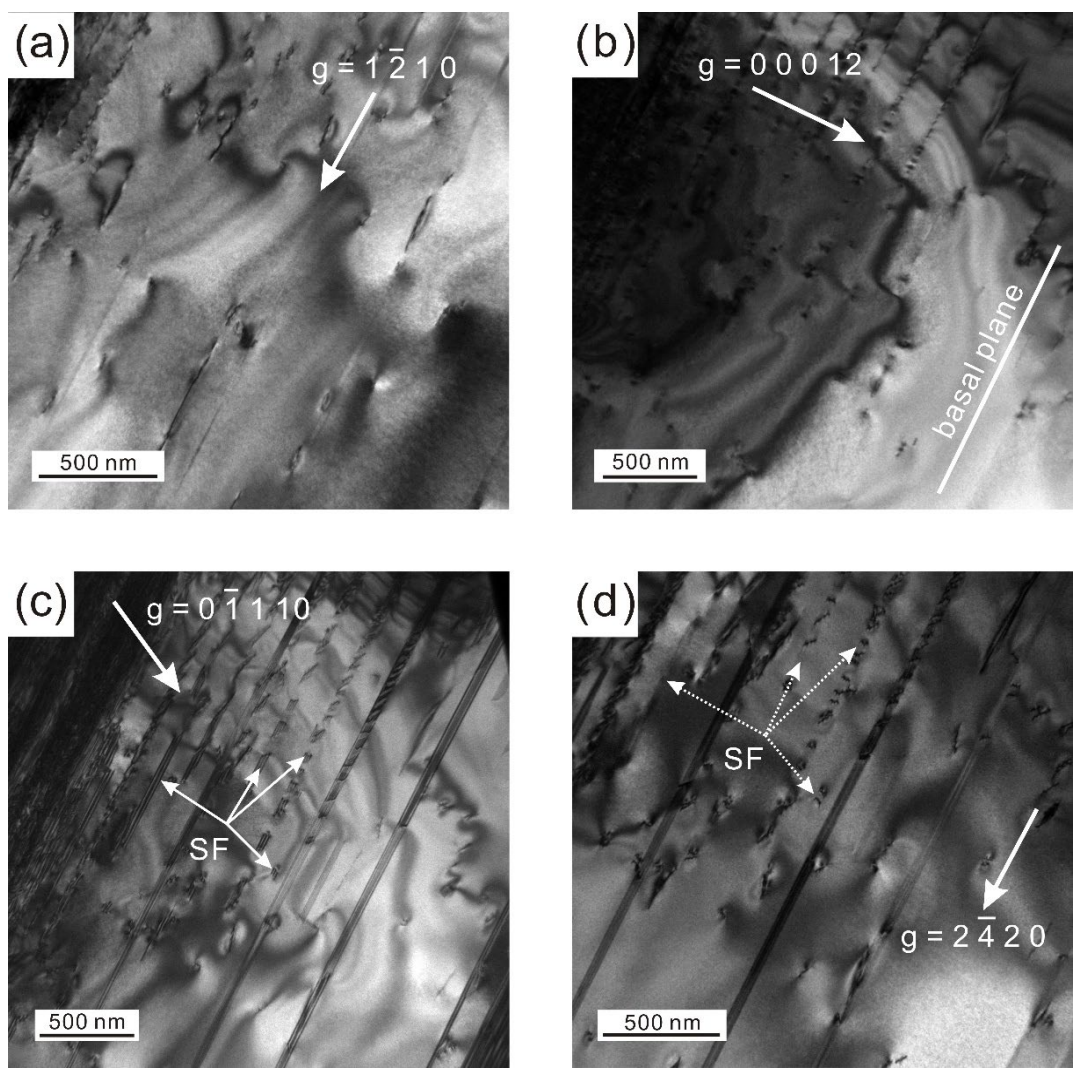

*Fig. 8 TEM bright-field images of a Co-54 Nb micropillar with 46º[0001] loading direction (oriented for $(0001)\langle 1\bar{1}00\rangle$ slip) taken along $[\bar{1}010]$ zone axis using (a) $g = 1\bar{2}10$ and (b) $g = 00012$, and along $[\overline{10}010\bar{1}]$ zone axis using (c) $g = 0\bar{1}110$ and (d) $g = 2\bar{4}20$.*

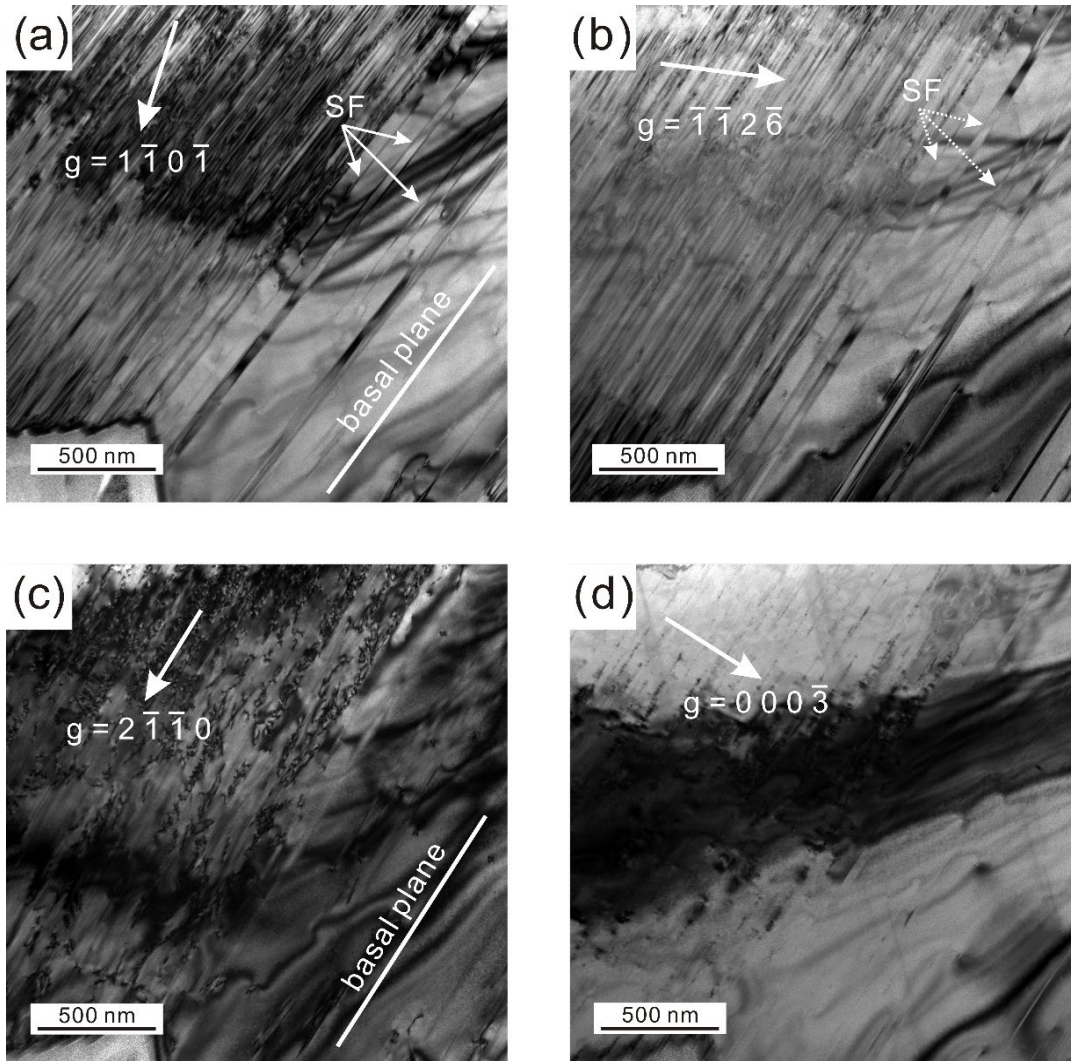

*Fig. 9 TEM bright-field images of a Co-54Nb micropillar with 47º[0001] loading direction (oriented for $(0001)\langle11\overline{2}0\rangle$ slip) taken along $[11\overline{2}0]$ zone axis using (a) $g = 1\overline{1}0\overline{1}$, along $[02\overline{2}\overline{1}]$ zone axis using (b) $g = \overline{1}\,\overline{1}2\overline{6}$, and along $[01\overline{1}0]$ zone axis using (c) $g = 2\overline{1}\,\overline{1}0$ and (d) $g = 000\overline{3}$.*

TEM bright-field images of a Co-54Nb micropillar with 47º[0001] loading direction (oriented for $(0001)\langle11\overline{2}0\rangle$ slip) taken under different two-beam conditions are shown in Figs. 9 (a-d). A large number of stacking faults on the basal plane were observed when imaged along the $[11\overline{2}0]$ zone axis using $g = 1\overline{1}0\overline{1}$ (Fig. 9 (a)). They are out of contrast when imaged along the $[02\overline{2}\overline{1}]$ zone axis using $g = \overline{1}\,\overline{1}2\overline{6}$ (Fig. 9 (b)). When imaged along the $[01\overline{1}0]$ zone axis using $g = 2\overline{1}\,\overline{1}0$ (Fig. 9 (c)), the stacking faults on the basal slip plane are out of contrast but arrays of partial dislocations were clearly observed. All of the dislocations and stacking faults are out of contrast using $g = 000\bar{3}$ (Fig. 9 (d)). A high density of $\frac{1}{3}\langle 1\bar{1}00 \rangle$ partial dislocations bounding stacking faults on the basal plane was observed in both the Co-54Nb micropillars oriented for $(0001)\langle 1\bar{1}00 \rangle$ and those oriented for $(0001)\langle 11\bar{2}0 \rangle$ slip. In both orientations, the basal slip traces transect the entire micropillar and have slip offsets of hundreds of nanometres, which indicates that the deformation of the Co-54Nb micropillars is controlled by repeated emission and subsequent glide of partial dislocations regardless of orientation.

### 3.3.3   Co-49Nb micropillars deformed at 600 ⁰C

TEM images of a Co-49Nb micropillar with 34⁰ [0001] orientation compressed at 600 ⁰C were taken under different two-beam conditions (Figs. 10 (a-c)). The pillar top was heavily deformed and shows a high density of defects lying on the basal plane, which is consistent with the dense basal slip traces at the pillar top (Fig. 5 (b)). The defect density decreases and individual defects can be clear seen in the middle of the micropillar. As shown in the enlarged images (Figs. 10 (d-f)) of the yellow squares, arrays of partial dislocations bounding stacking faults are parallel to the basal plane. The contrast of stacking faults was clearly seen when imaged along the $[\bar{1}\bar{0}\,0\,10\,\bar{1}]$ zone axis using $g = 0\,\bar{1}110$ (Fig. 10 (d)). When imaged along the $[\bar{1}010]$ zone axis using $g = \bar{1}2\bar{1}0$ (Fig. 10 (e)), the partial dislocations are visible but the stacking faults are out of contrast. All of the defects imaged along the $[\bar{1}010]$ zone axis with $g = 000\bar{3}$ (Fig. 10 (f)) are out of contrast. Therefore, in the Co-49Nb micropillars the defects induced by compression at 600 ⁰C are $\frac{1}{3}\langle 1\bar{1}00 \rangle$ partial dislocations bounding stacking faults, which indicates that the plastic deformation of the Co-49Nb micropillars at 600 ⁰C occurs via glide of partial dislocations on the basal plane.

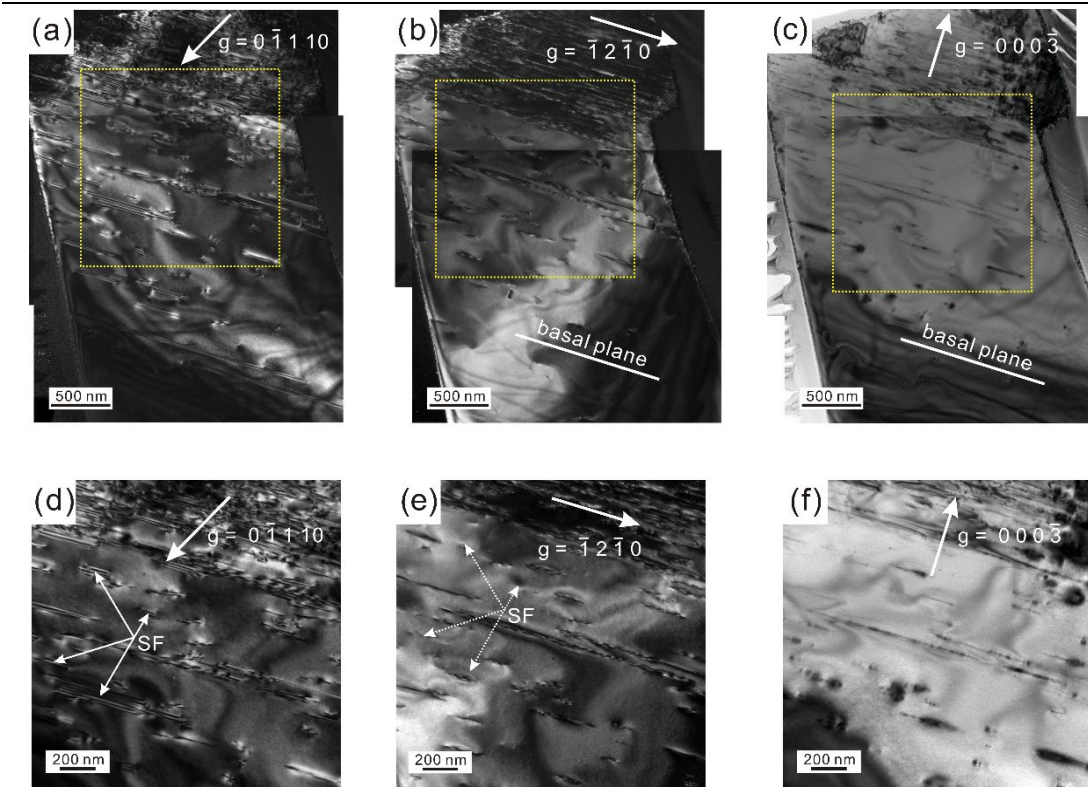

*Fig. 10 TEM dark-field images of a Co-49Nb micropillar with 34º[0001] orientation deformed at 600 ºC taken (a) along $[\overline{1}0\,0\,10\,\overline{1}]$ zone axis using $g = 0\,\overline{1}110$ and (b) along $[\overline{1}010]$ zone axis using $g = \overline{1}2\,\overline{1}0$. (c) TEM bright-field image taken along $[\overline{1}010]$ zone axis with $g = 000\overline{3}$. The enlarged images of the areas marked by yellow squares in (a-c) are shown in (d-f), respectively.*

### 3.4 Critical resolved shear stress

The slip traces and TEM analyses confirm that at room temperature, dislocation slip in the μ-$Co_7Nb_6$ phase is found exclusively on the basal plane. While the Co-49Nb micropillars deform via glide of $\frac{1}{3}\langle 11\overline{2}0\rangle$ full dislocations on the basal plane, the Co-54Nb micropillars deform via glide of $\frac{1}{3}\langle 1\,\overline{1}00\rangle$ partial dislocations on the basal plane. Therefore, the highest Schmid factors for the primary $(0001)\langle 11\overline{2}0\rangle$ and $(0001)\langle 1\,\overline{1}00\rangle$ slip systems are used to calculate the CRSS for basal slip of the Co-49Nb and Co-54Nb micropillars deformed at room temperature, respectively. Similarly, the highest Schmid factor for the primary $(0001)\langle 1\,\overline{1}00\rangle$ slip system is used to calculate the CRSS for basal slip of the Co-49Nb micropillars deformed at 600 °C.
The CRSS of the micropillars of the μ-Co$_7$Nb$_6$ phase with different sizes is shown in
Fig. 11 as a function of composition. These micropillars are oriented for basal slip. The
CRSS of the Co-49Nb micropillars of 0.8, 2 and 3 μm diameter deformed at room
temperature is 3.8 ± 0.8, 2.3 ± 0.9 and 1.8 ± 0.4 GPa, respectively. The CRSS of the
Co-49Nb micropillars of 2 μm diameter deformed at 600 °C is 0.8 ± 0.1 GPa, which is
significantly lower than that of the Co-49Nb micropillars deformed at room temperature.
The respective CRSS of the Co-52Nb micropillars of 0.8 and 2 μm diameter is 3.2 ±
0.7 and 2.5 ± 0.9 GPa, assuming full dislocation slip is dominant. The CRSS of the Co-
49Nb and Co-52Nb micropillars of the same size are very close and the differences
between them are within error bars. However, the CRSS of the Co-54Nb micropillars
is significantly lower than that of the Co-49Nb and Co-52Nb micropillars with the same
pillar size. The CRSS of the Co-54Nb micropillars having 0.8, 2, 3 and 5 μm diameter
is 1.0 ± 0.3 GPa, 0.7 ± 0.2 GPa, 0.5 ± 0.2 GPa and 0.46 ± 0.08 GPa, respectively.

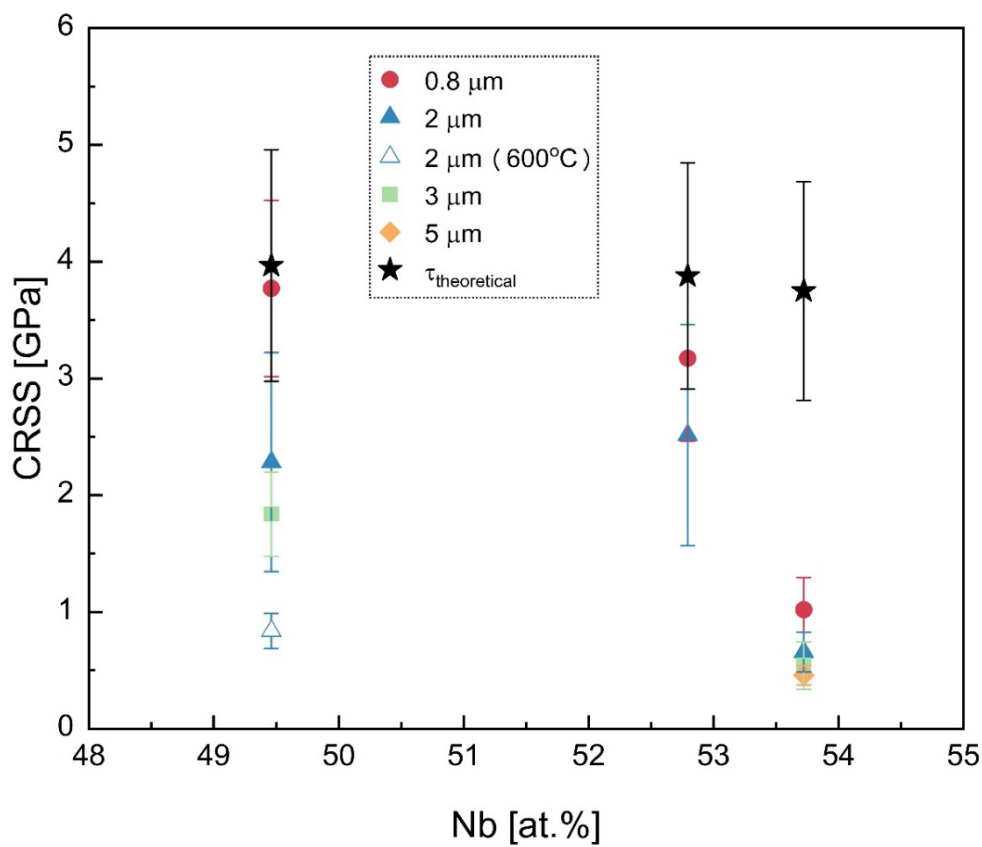

*Fig. 11 The experimental CRSS values of the μ-Co$_7$Nb$_6$ phase micropillars of different
sizes as well as the theoretical shear stress for dislocation nucleation as a function
of composition. The error bars for the experimental CRSS values represent their
standard deviations. For the theoretical stress, the error bars span between the values
of a quarter and half dislocation loop [29].*

## 4    Discussion

### 4.1   Full versus partial dislocation slip

#### 4.1.1   Full dislocation slip in Co-49Nb micropillars at room temperature

As shown in Fig. 1, the μ-$Co_7Nb_6$ phase consists of building blocks of the $MgCu_2$-type Laves layers and the $Zr_4Al_3$ layers. While no information of plasticity of the $Zr_4Al_3$ phase has been reported, there are two possible mechanisms of plastic deformation on the close-packed plane of the Laves phase ((0001) plane for the hexagonal C14 and C36 Laves phases and $\{1\overline{1}1\}$ plane for the cubic C15 Laves phase). The most widely assumed deformation mechanism of the Laves phase, which has been confirmed by experiments [30] and simulations [31, 32], is the synchroshear mechanism. It operates within the triple layer structural unit and consists of two simultaneous shear deformations in different directions on adjacent atomic layers [33]. The undulating slip mechanism proposed in [34] is actually identical to the synchroshear mechanism [32]. The other possible deformation mechanism is crystallographic slip [31, 32] operating between the triple layer and the kagomé layer. Density function theory (DFT) calculations and atomistic simulations show that the energy barrier for the crystallographic slip is higher than that for synchroshear in the C15 $Cr_2Nb$ and $Cr_2Ta$ [31], C14 $Cr_2Nb$ [34] and C14 $Mg_2Ca$ [32] Laves phases. The local minimum in the energy profile of the crystallographic slip along the partial dislocation glide path is very shallow [31], which indicates that dislocation dissociation is unlikely. As the crystal structure changes from the $Co_2Nb$ Laves phase to the μ-$Co_7Nb_6$ phase, the interplanar spacing between the triple layer and the kagomé layer $d_{triple-kagome}$ increases but the interplanar spacing within the triple layer $d_{triple}$ slightly decreases. In the Laves phase the ratio $d_{triple-kagome}^{Laves}/d_{triple}^{Laves}$ is 3, whereas in the μ phase the ratio $d_{triple-kagome}^{\mu}/d_{triple}^{\mu}$ is about 5. From a geometric point of view, while the energy barrier for synchroshear is likely to increase, the energy barrier for the crystallographic slip between the triple layer and the kagomé layer is expected to decrease as the structure changes from the Laves phase to the μ phase.

In the μ-$Co_7Nb_6$ phase (Fig. 1), besides the basal plane within the triple layer $P_{triple}$ and the basal plane between the triple layer and the kagomé layer $P_{triple-kagomé}$, there are two additional possible basal slip planes $P_{kagomé-CN14}$ (the basal plane between the CN 14 layer and the kagomé layer) and $P_{CN14-CN15}$ (the basal plane between the CN14 layer and CN15 layer). According to the Peierls-Nabarro model [35, 36], dislocation slip is most likely to operate on the $P_{triple-kagomé}$ plane due to the lowest ratio of the interplanar spacing to the Burgers vector, $d/b$. However, the Peierls-Nabarro model is perhaps not directly applicable to the synchroshear mechanism where the dislocation cores of the synchro-Shockley partial dislocation are not co-planar and spread on two adjacent planes [36]. Moreover, estimation of the lattice resistance of different slip systems on the basis of $d/b$ works well in BCC, HCP and FCC metals, but becomes more difficult in complex crystals, which are formed by dense packing of coordination polyhedral [27] .

We therefore apply here the method suggested by Pal et al. [27] based on the degree of geometric overlap during slip to estimate the most plausible basal slip plane of crystallographic slip in the μ-$Co_7Nb_6$ phase. The geometric γ-surfaces (g-surfaces) based on the topological overlap of atomic volume associated with the relative rigid body motion across the possible basal planes (Fig. 1) were calculated. Although the atomic interactions were not considered in the calculations, the g-surfaces can provide a qualitative indication of the most favourable slip system in the μ-$Co_7Nb_6$ phase as the high strength of the μ-$Co_7Nb_6$ phase is not associated with strong directional bonds but rather with the complexity of the crystal structure [37]. The g-surfaces for $P_{CN14-CN15}$, $P_{kagomé-CN14}$, $P_{triple-kagomé}$, and $P_{triple}$ basal planes of the μ-$Co_7Nb_6$ phase are shown in Figs. 12 (a-d), respectively. While there is a shallow minimum on the g-surface for $P_{kagomé-CN14}$ (Fig. 12 (b)) along $\frac{1}{3}\langle 1\bar{1}00\rangle$ direction, there are no minima on the g-surfaces for the other basal planes along either $\frac{1}{3}\langle 11\bar{2}0\rangle$ or $\frac{1}{3}\langle 1\bar{1}00\rangle$ direction, which indicates that in the μ-$Co_7Nb_6$ phase dislocation dissociation is unfavourable for crystallographic slip regardless of the basal slip plane. The g-surfaces for $P_{triple-kagomé}$ (Fig. 12 (e)), and $P_{triple}$ (Fig. 12 (f)) of the C15-$Co_2Nb$ Laves phase were also calculated for comparison. While there is a shallow minimum on the g-surface for $P_{triple-kagomé}$ (Fig. 12 (e)) along $\frac{1}{3}\langle 1\bar{1}00\rangle$ direction, there are no minima on the g-surfaces for $P_{triple}$ (Fig. 12 (f)) along either $\frac{1}{3}\langle 11\bar{2}0\rangle$ or $\frac{1}{3}\langle 1\bar{1}00\rangle$ direction, which suggests dislocation dissociation is also unfavourable for crystallographic slip in the C15-$Co_2Nb$ Laves phase.

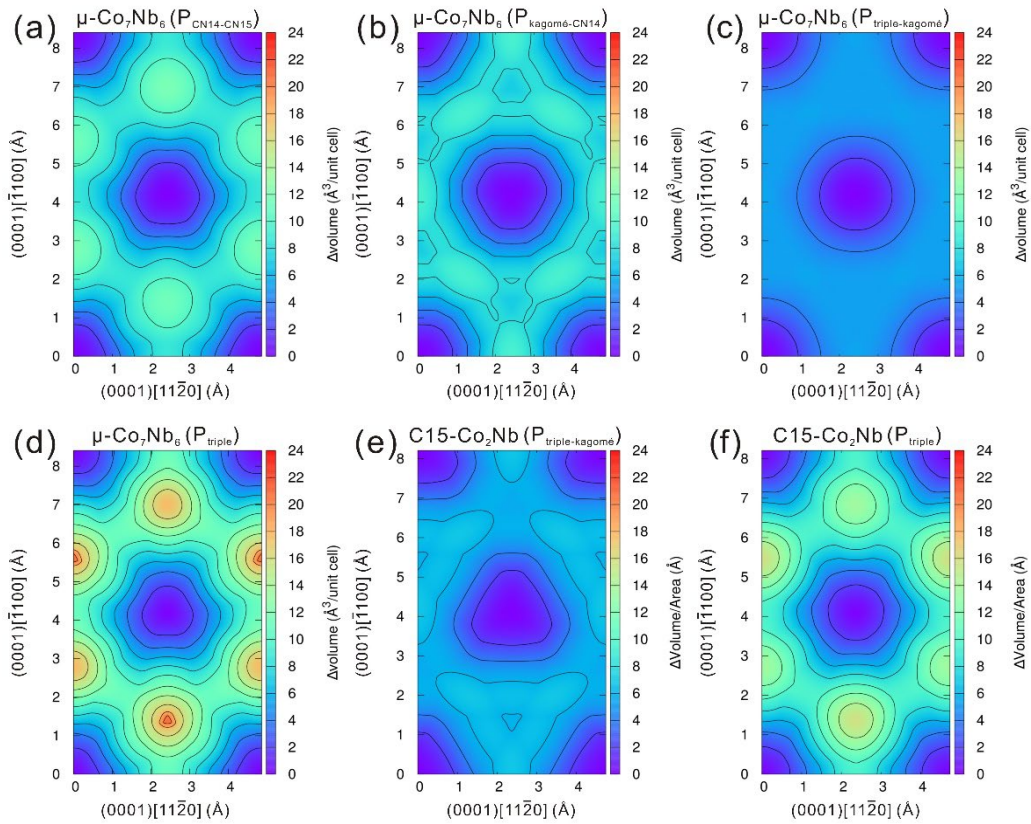

*Fig. 12 The g-surfaces for (a) $P_{CN14-CN15}$, (b) $P_{kagomé-CN14}$, (c) $P_{triple-kagomé}$, and (c) $P_{triple}$ basal planes of the µ-Co$_7$Nb$_6$ phase as well as (e) $P_{triple-kagomé}$, and (f) $P_{triple}$ basal planes of the C15-Co$_2$Nb phase. There are no pronounced local minima on the g-surfaces for different basal planes along either $\frac{1}{3}\langle 11\bar{2}0 \rangle$ or $\frac{1}{3}\langle 1\bar{1}00 \rangle$ direction of both the µ-Co$_7$Nb$_6$ phase and the C15-Co$_2$Nb Laves phase, indicating that dislocation dissociation is unfavourable for crystallographic slip regardless of the basal slip plane.*

The projections of the g-surfaces for different basal planes along $\frac{1}{3}\langle 1\bar{1}00 \rangle$ and $\frac{1}{3}\langle 11\bar{2}0 \rangle$ direction in the µ-Co$_7$Nb$_6$ phase are shown in Figs. 13 (a-d), respectively. The projection of the g-surface for P$_{triple-kagomé}$ along either $\frac{1}{3}\langle 11\bar{2}0 \rangle$ (Fig. 13 (a)) or $\frac{1}{3}\langle 1\bar{1}00 \rangle$ (Fig. 13 (b)) direction shows the lowest overlapped volume, indicating that P$_{triple-kagomé}$ is the most plausible basal plane for crystallographic slip. The absence of local minimum on g-surface for P$_{triple-kagomé}$ along $\frac{1}{3}\langle 1\bar{1}00 \rangle$ (Fig. 13 (b)) direction suggests that crystallographic slip on P$_{triple-kagomé}$ is likely to occur via full dislocation slip without dislocation dissociation. In the C15-Co$_2$Nb Laves phase, the projection of the g-surface for $P_{triple-kagomé}$ shows a lower overlapped volume along either $\frac{1}{2}\langle 110 \rangle$ (Fig. 13 (c)) or $\frac{1}{6}\langle 11\overline{2} \rangle$ (Fig. 13 (d)) direction than that for $P_{triple}$, which suggests that in the C15-Co$_2$Nb Laves phase crystallographic slip also tends to occur between the triple layer and the kagomé layer. The shallow local minimum in the projection of the g-surface for $P_{triple-kagomé}$ along $\frac{1}{6}\langle 11\overline{2} \rangle$ (Fig. 13 (d)) agrees with the DFT calculations [31], which indicates that dislocation dissociation is unfavourable for crystallographic slip in the C15-Co$_2$Nb.

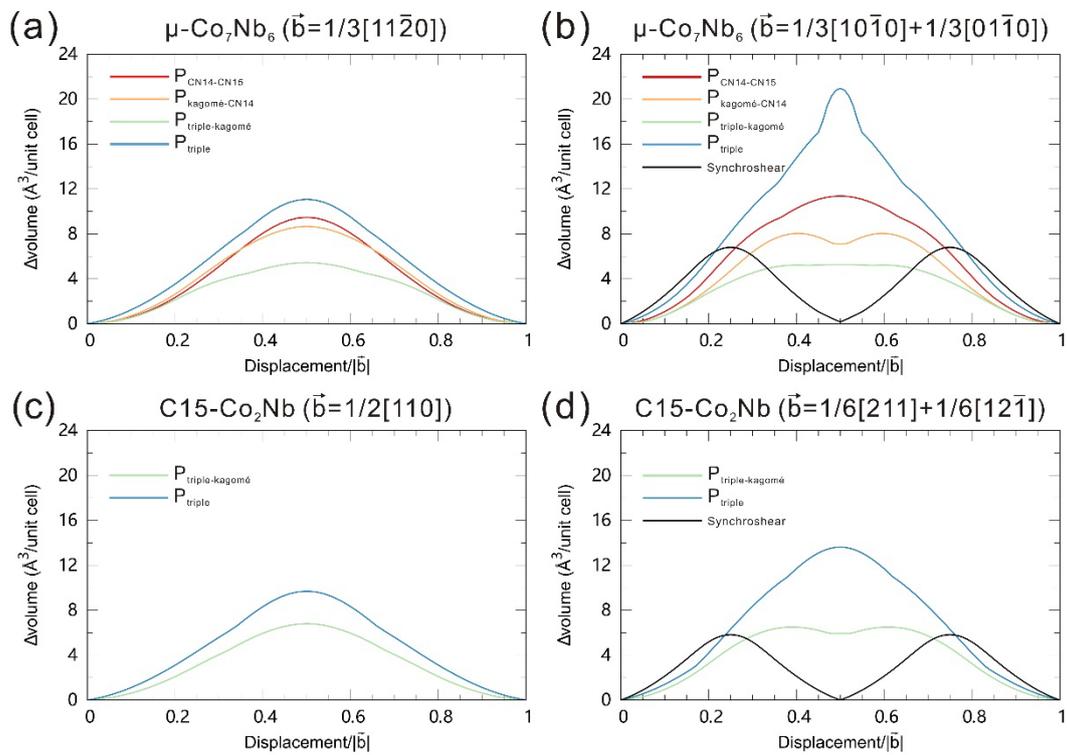

*Fig. 13 The projections of the g-surfaces of crystallographic slip on the $P_{CN14-CN15}$, $P_{Kagomé-CN14}$, $P_{triple-kagomé}$, and $P_{triple}$ basal planes and synchroshear in the μ-Co$_7$Nb$_6$ phase along (a) full ($\frac{1}{3}[11\overline{2}0]$) and (b) partial ($\frac{1}{3}[10\overline{1}0]+\frac{1}{3}[01\overline{1}0]$) slip direction and in the C15-Co$_2$Nb crystals along (c) full ($\frac{1}{2}[110]$) and (d) partial ($\frac{1}{6}[211]+\frac{1}{6}[12\overline{1}]$) slip direction.*

The overlapped volume caused by synchroshear was calculated and compared with that caused by crystallographic slip in both the μ-Co$_7$Nb$_6$ phase (Fig. 13 (b)) and the C15-Co$_2$Nb Laves phase (Fig. 13 (d)). While the overlapped volume during crystallographic slip is calculated by moving the atoms in the upper half crystal above a basal slip plane along a specific direction with respect to those in the bottom half crystal, the overlapped volume during synchroshear is calculated by simultaneously shifting the two adjacent atomic layers within a triple layer along the two partial dislocation slip directions. In the C15-Co$_2$Nb Laves phase, the maximum overlapped volume caused by synchroshear is lower than that caused by crystallographic slip on P$_{triple-kagomé}$ along $\frac{1}{6}\langle 11\overline{2} \rangle$ direction, which agrees with the results of DFT calculations for the NbCr$_2$ and TaCr$_2$ Laves phases [31]. However, in the μ-Co$_7$Nb$_6$ phase, the maximum overlapped volume caused by synchroshear is higher than that caused by crystallographic slip on P$_{triple-kagomé}$ along $\frac{1}{3}\langle 1\overline{1}00 \rangle$ direction, indicating that in the μ-Co$_7$Nb$_6$ phase crystallographic slip on P$_{triple-kagomé}$ is geometrically more favourable than synchroshear presumably due to the increased $d^{\mu}_{triple-kagome}/d^{\mu}_{triple}$ ratio.

The most favourable slip mechanism of the μ-Co$_7$Nb$_6$ phase predicted by the g-surfaces is consistent with the results of slip trace analyses and TEM investigations of the Co-49Nb micropillars deformed at room temperature. Even though the Co-49Nb micropillars are in orientations where dislocation dissociation is promoted during compression due to the large difference between the Schmid factors of leading partial dislocations ($m_L = 0.5$) and trailing partial dislocations ($m_T = 0.25$), full dislocation slip on the basal plane without dissociation was observed at room temperature.

### 4.1.2 Partial dislocation slip in Co-54Nb micropillars at room temperature

The stoichiometric composition of the Fe$_7$W$_6$-type μ-phase is 46.2 at.% Nb. However, the μ-Co$_7$Nb$_6$ phase has an extended homogeneity range of 46.5 - 56.1 at.% Nb on the Nb-rich side of the stoichiometric composition. It has been reported that the excess Nb atoms in the μ-Co$_7$Nb$_6$ phase tend to replace the Co atoms in the triple layers and at 51.5 at.% Nb, 70 % of the Co atoms in the triple layers are replaced by Nb atoms [38]. Assuming the excess Nb atoms in the μ-Co$_7$Nb$_6$ phase in the composition range of 49.5 – 53.7 at.% Nb substitute the Co atoms in the triple layers, 43.5 % and 98.1 % of Co atoms in the triple layers are replaced by Nb atoms at 49.5 and 53.7 at.% Nb, respectively. It indicates that at 53.7 at.% Nb, nearly all of the Co atoms in the triple layers are replaced by Nb atoms, and thus the triple layers may become slabs of pure Nb. Since the atomic environment of the triple layer of pure Nb resembles that of elemental Nb with likely more metallic bonding, the bond strength of the triple layers might be weakened, and thus the MgCu$_2$-type Laves layer may become a compliant layer with respect to the Zr$_4$Al$_3$ layer in the μ-Co$_7$Nb$_6$ phase at 53.7 at.% Nb. As the shear modulus of Nb ($G_{Nb}$ = 30.9 GPa [39]) is relatively low, in the triple layers consisting of pure Nb the critical stress of dislocation nucleation and the Peierls stress are both expected to decrease. Therefore, basal slip is likely to occur more easily on the basal planes in the $MgCu_2$-type Laves layers due to the reduction in the local shear modulus. Similar softening behaviour caused by inhomogeneous elasticity has been reported in the MAX phase with a layered structure [40]. The differences in electronegativity between the adjacent layers give rise to non-uniform elastic deformation of the unit cell and changes in $d/b$. As the electronegativity difference between the adjacent layers increases, the shear modulus of the complaint layers decreases with respect to that of the stiff layers and the dislocation width increases due to the reduction of the misfit energies, leading to a dramatic reduction in the Peierls stress [40].

Moreover, in contrast to the Co-49Nb micropillars, all of the Co-54Nb micropillars, even those oriented for full dislocation slip on the basal plane show high densities of partial dislocations bounding stacking faults on the basal plane after compression. This indicates that a slight change of composition from 49.5 to 53.7 at.% Nb not only dramatically influences the mechanical properties of the $\mu$-$Co_7Nb_6$ phase but also changes the basal slip mechanism. Atomic-resolution electron microscopy studies beyond the scope of this manuscript would be needed to reveal the atomic configurations of the triple layers and the partial dislocations created in the $\mu$-$Co_7Nb_6$ phase at 53.7 at.% Nb.

From a geometrical point of view, crystallographic slip between the triple layer and the kagomé layer is favourable in the $\mu$-$Co_7Nb_6$ phase. However, the geometrical predication using the overlapped atomic volume excludes the atomic interactions as well as the contribution of structural rearrangement such as short-range atomic shuffling [32, 41] and in-plane lattice relaxation [42] on overcoming the energy barrier to slip events. Besides, the constitutional defects in the $\mu$-$Co_7Nb_6$ phase at 53.7 at.% Nb might change dramatically the atomic configurations and the atomic bonds. In this case, the deformation mechanisms and the corresponding energy barriers estimated by the geometric overlap method may not be reliable.

### 4.1.3   Partial dislocation slip in Co-49Nb micropillars at 600 ºC

As the temperature increases from room temperature to 600 ºC, the deformation behaviour of the Co-49Nb micropillars oriented for basal slip changes from elastic deformation followed by an abrupt strain burst at room temperature to stable and continuous plasticity and, moreover, the average CRSS of basal slip decreases dramatically from $2.3 \pm 0.9$ to $0.8 \pm 0.1$ GPa, which indicates that basal slip of the Co-49Nb micropillars is thermally activated and therefore easier at elevated temperature. In contrast, the Co-49Nb micropillars oriented for non-basal slip show high yield stresses and large strain bursts at both room temperature and 600 °C. It indicates that the dramatic changes in the deformation behaviour and CRSS values of the Co-49Nb micropillars oriented for basal slip at 600 °C are likely related to the transition in basal slip mechanism facilitated by thermal activation rather than the decrease in elastic constants with temperature. While full dislocation slip on the basal plane is dominant in the Co-49Nb micropillars at room temperature, the presence of high densities of partial dislocations bounding stacking faults on the basal plane indicates that basal slip in the Co-49Nb micropillars mainly occurs via a partial dislocation slip mechanism at 600 °C. Although crystallographic slip is geometrically favourable in the $\mu$-Co$_7$Nb$_6$ phase at room temperature, the absence of pronounced local minima on g-surfaces for crystallographic slip on different basal planes of the $\mu$-Co$_7$Nb$_6$ phase suggests that the partial dislocation slip on the basal plane in the Co-49Nb micropillars at 600 °C is unlikely to occur via crystallographic slip. Since synchroshear within the tightly packed triple layers is a thermally activated process, it can be activated when the crystal structure is substantially softened and the self-pinning resistance is overcome at elevated temperature [43, 44]. It has been confirmed by atomistic simulations [45] that the two synchro-Shockley dislocations associated with synchroshear propagate via a kink-pair propagation and a non-sequential shuffling mechanism, respectively and their motions are both thermally activated. At elevated temperature the thermal fluctuation can help to overcome the nucleation and migration barriers, increase the mobility of the synchro-Shockley dislocations, and eventually reduce the critical stress of nucleation and propagation [45]. How exactly the partial dislocation slip is thermally activated and where the transition occurs in the $\mu$-Co$_7$Nb$_6$ phase will require careful further work at smaller temperature intervals and dedicated simulations of the dynamics of both mechanisms.

## 4.2   Size effect on the CRSS

The respective theoretical shear stresses of circular micropillars for nucleation of a partial and full dislocation loop are estimated to be $\tau_{t,partial} = \dfrac{cGb_{partial}}{2\pi e^2 \rho}\left(\dfrac{2-\nu}{1-\nu}\right) + \dfrac{c\gamma}{b}$

[46] and $\tau_{t,full} = \dfrac{cGb_{full}}{2\pi e^2 \rho}\left(\dfrac{2-\nu}{1-\nu}\right)$ [29], where $G$ is the shear modulus, $b_{partial}$ and $b_{full}$ are

the respective Burgers vectors of partial and full dislocation, $\nu$ is the Poisson's ratio, $\rho$ is the linear cutoff parameter estimated to be $b/2$ [47], $c$ is a correction factor ranging from 0.3 (quarter dislocation loop) to 0.5 (half dislocation loop) [29], and $\gamma$ is the stacking fault energy. The Poisson's ratio of the $\mu$-Co$_7$Nb$_6$ phase is taken to be 0.31 [37].

Compared to the first term representing the elastic energy contribution in $\tau_{t,\,partial}$, the term of stacking fault energy is negligible [46]. Therefore, the theoretical shear stress of a micropillar for dislocation nucleation is estimated to be $0.03\,G - 0.05\,G$, depending on the size of the dislocation loop. The average shear modulus of the Co-49Nb, Co-52Nb and Co-54Nb alloys estimated from nanoindentation tests is 94, 92 and 89 GPa, respectively. The CRSS (Fig. 11) of the Co-49Nb and Co-52Nb micropillars of 0.8 μm diameter is close to the theoretical shear stress but the CRSS decreases significantly as pillar size increases. At the base of the deformed Co-49Nb micropillars (supplementary Fig. 1), which is under a lower stress due to the pillar taper and far away from the heavily deformed pillar top, dislocations are rarely seen. It indicates that the Co-49Nb micropillars, especially the small micropillars of 0.8 μm diameter, may have a lack of pre-existing dislocation sources. Therefore, the deformation of the Co-49Nb micropillars is likely controlled by dislocation nucleation and the subsequent dislocation avalanches. As pillar size increases, the probability to find pre-existing dislocation sources or favourable dislocation nucleation sites increases, and thus the average CRSS decreases. In the Co-54Nb micropillars (supplementary Fig. 2), besides the slip bands consisting of high densities of dislocations there are individual dislocations at the pillar base which is far away from the deformed region. It indicates that there might be pre-existing dislocations which could be activated during loading in the Co-54Nb micropillars. As pillar size increases, the number of pre-existing dislocation sources increases, and thus the average CRSS of the Co-54Nb micropillars decreases.

The effect of pillar size on the CRSS is described by a power-law relation $(\tau_{CRSS}/G) = A \cdot (D/b)^n$ [48], where $A$ is a fitting parameter, $D$ is the top diameter of the micropillar and $n$ is the power-law exponent. The Burgers vectors $b_{full} = 0.4898$ nm and $b_{partial} = 0.2828$ nm [23] are used for the calculations of Co-49Nb and Co-54Nb micropillars, respectively. As shown in the Fig. 14, the data were fitted with the power-law relation and the power-law exponents $n$ for the Co-49Nb and Co-54Nb micropillars are 0.58 and 0.51, respectively. The $n$ values of the micropillars of the μ-$Co_7Nb_6$ phase are rather high for a hard phase, in comparison with the hexagonal and cubic $NbCo_2$ Laves phases [46] and other hard materials [21, 22]. Why the μ-$Co_7Nb_6$ phase shows a more pronounced size effect on the CRSS than the closely related $NbCo_2$ Laves phase is not yet clear.

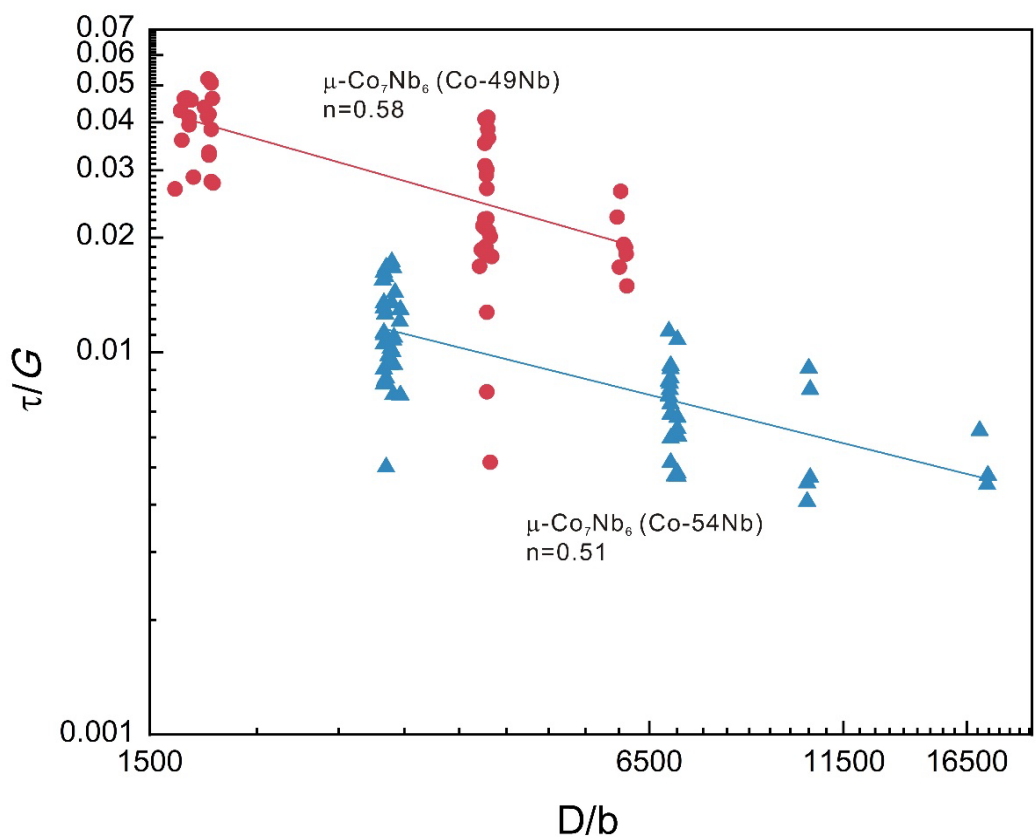

*Fig. 14 The size scaling of the CRSS of the Co-49Nb and Co-54Nb micropillars. Their respective power law exponents n are 0.58 and 0.51.*

## 5   Conclusions

The influences of composition and temperature on the plasticity of the μ-$Co_7Nb_6$ phase were studied by micropillar compression tests and the defects induced by deformation were investigated by TEM. The present work shows that the mechanical properties and deformation mechanism of the μ-$Co_7Nb_6$ phase can be dramatically influenced by the composition and temperature. The main conclusions are as follows:

(1) The micropillars of the μ-$Co_7Nb_6$ phase mainly deform by basal slip without showing clear evidence of non-basal slip. While Co-49Nb and Co-52Nb micropillars show high yield stresses and an abrupt large strain burst at the onset of yielding regardless of orientation, the Co-54Nb micropillars oriented for basal slip yield at much lower stresses and show intermittent small strain bursts during plastic deformation. The CRSS of the micropillars is very similar at 49.5 and 52.8 at.% Nb but decreases for 53.7 at.% Nb.

(2) TEM investigations show that while the Co-49Nb micropillars oriented for basal slip mainly deform by $\frac{1}{3}\langle 11\bar{2}0 \rangle$ full dislocation slip on the basal plane without dislocation dissociation at room temperature, the Co-54Nb micropillars oriented for basal slip mainly deform by $\frac{1}{3}\langle 1\bar{1}00\rangle$ partial dislocation slip on the basal plane.

(3) The geometric γ-surfaces for different basal planes of the μ-Co$_7$Nb$_6$ phase indicate that full dislocation slip between the triple layer and the kagomé layer is geometrically favourable at room temperature due to the increased $d^{\mu}_{triple-kagome} \big/ d^{\mu}_{triple}$ ratio in the μ-Co$_7$Nb$_6$ phase.

(4) At 600 °C the Co-49Nb micropillars oriented for basal slip show stable and continuous plasticity and a transition of basal slip mechanism from full to partial dislocation slip accompanied by a strong reduction in the CRSS.

## 6 Acknowledgement

The authors thank Dunming Wu, David Beckers and Arndt Ziemons for their help in sample preparations, Martin Heller and Risheng Pei for their help in FIB milling, James Best for his help in nanoindentation and micropillar compression tests, as well as Siyuan Zhang, Frank Stein and Stefanie Sandlöbes-Haut for fruitful discussions. Funding by the Deutsche Forschungsgemeinschaft (DFG) in project KO 4603/2-2 and SFB1394 (project ID 437514011) is gratefully acknowledged. This project has received funding from the European Research Council (ERC) under the European Union's Horizon 2020 Research and Innovation Programme (Grant Agreement No. 852096 FunBlocks).